\documentstyle[12pt]{article}
\begin{document}

\title{
\begin{flushright}
{\normalsize VPI-IPPAP-98-10}\\ \ \\
\end{flushright}
Probing Spontaneous CP-Violation in
Supersymmetric Models via B-Decays}
\author{ Oleg Lebedev\\
     \it{Virginia Polytechnic Insitute and State University}\\
     \it{Department of Physics}\\
     \it{Institute for Particle Physics and Astrophysics}\\
     \it{Blacksburg, Virginia 24061}}
\maketitle
\thispagestyle{empty}
\begin{abstract}
We study implications of spontaneous CP-violation in minimal susy
models for CP-asymmetries in rare $B,\bar B$-decays.
In particular, we estimate
characteristic values of the angles of the unitarity triangle
and show that sizeable
CP-violating effects result from $B-\bar B$ mixing only.
Significant deviations from the SM predictions are pointed out.
\end{abstract}
\newpage
\section{Introduction}
One of the most intriguing open questions in particle physics
is understanding the origin of CP-violation.
Even though the observable CP-violating effects in kaon decays
can be accommodated within the Standard Model via the
Cabibbo-Kobayashi-Maskawa (CKM) mechanism [1], the real nature of
CP-violation
is still to be uncovered. 

A careful study of CP-violating effects in B-decays would reveal whether the
CKM model provides an adequate description of CP-violation in nature.
In particular, the CKM model implies the existence of a nondegenerate
unitarity triangle [2]: the condition of orthogonality of the columns or rows
of a unitary matrix can be represented by a triangle in the complex plane.
The angles of this triangle correspond to the relative complex phases
between the CKM matrix elements. For example, since sizeable complex phases
are
expected in the CKM matrix elements involving the first and third generations,
a triangle formed by $V(ub),\;V^{*}(td)$, and $-V(cb)\sin \theta_{c}$
should have relatively large angles: their typical values range from
$15^{\circ}$ to $145^{\circ}$ (see, for example, [3]).
Independent measurements of the sides and angles of this triangle would
overdetermine the unitarity triangle allowing us to check explicitly
the validity of the CKM approach. Any appreciable deviation from
the Standard
Model predictions would indicate the presence of ``New Physics"
such as the existence of a fourth family, supersymmetry, etc.

In this letter we study implications of an alternative to the
CKM approach - a model in which CP-symmetry is broken spontaneously.
In particular, we will consider minimal susy models and concentrate on
CP-violating effects in B-decays which allow us to extract information
about the angles of the unitarity triangle. 

\section{Spontaneous CP-violation in SUSY Models}
The possibility of spontaneous CP-violation (SCPV) in the minimal susy models
has recently drawn considerable attention [6-13]. The basic idea is that,
in a general two Higgs doublet model, the Higgs fields can acquire complex
VEV's if the scalar potential is not Peccei-Quinn (PQ) invariant [4].
Phenomenologically acceptable susy models present a fertile ground for
SCPV since they require the presence of at least two Higgs doublets.

In the minimal supersymmetric Standard Model (MSSM), the desired PQ-breaking
terms can be generated only via radiative corrections [6] and, as a
result of the Georgi-Pais theorem [5], such a scenario predicts
the existence
of a light axion. Consequently, even though SCPV is possible in the MSSM
in principle [8,9], it is ruled out by experimental constraints on the
mass of the lightest Higgs boson [7].

The next-to-minimal supersymmetric
Standard Model (NMSSM) has been shown to be free of this problem [10]
and is, at the moment, experimentally viable.
The implications of this model for observable CP-violating effects in
$K-\bar K$ systems were first
studied by Pomarol [11].
He has shown that, 
in a favorable region of the parametric
space, this scenario can predict correct values of $\epsilon$ and $\epsilon '$
while complying with the experimental bounds on the Neutron Electric Dipole
Moment (NEDM). Our more recent analysis [12] showed that the requirement
that the squarks be sufficiently heavy (300-400 GeV) allows one
to enlarge the available region of the parametric space. Yet, even in this
case some fine-tuning is required and 
the values of the CP-phases would have to be small: from 0.01 to 0.1.

The experimental information on CP-violation in $K-\bar K$ systems and
constraints on the NEDM cannot distinguish between the
CKM model and susy models with spontaneously broken CP. To discriminate
against one of them, we have to combine
these data with B-decay phenomenology. In what follows, we estimate
the characteristic values of the angles of the unitarity triangle
$\alpha_{i}$ in the context of
spontaneous CP-violation in the NMSSM (or similar models).
We will see that they are significantly different from their Standard Model
counterparts.

\section{CP-violation in B Decays}
Let us consider nonleptonic B decays into CP eigenstates. The three angles of
the unitarity triangle $\alpha_{1-3}${\footnote{We follow the
notation of Ref.[3]}}  can be probed, for example, via
the following decays 
\begin{eqnarray}
&& B_{d} \rightarrow \psi   K_{s} \;\;\;\;\sim \sin 2\alpha_1 \;,\\
&& B_{d} \rightarrow \pi^{+} \pi^{-} \;\;\sim \sin 2\alpha_2 \;,\\
&& B_{s} \rightarrow \rho^0  K_{s} \;\;\;\sim \sin 2\alpha_3 \;.
\end{eqnarray}
In these decays, CP-violation manifests itself as a deviation of the decay
rate from a pure exponent $e^{-\Gamma t}$. Since the CKM model predicts
enormous
(up to 30$\%$) CP-asymmetries [14], these decays offer excellent
opportunities for observing
large CP-violating effects.
Neglecting a small{\footnote {For the $B_{s}-\bar B_{s}$ system it is
not negligible. The corresponding CP-asymmetry can 
be read off from
the $ e^{-{1\over2}\Delta \Gamma t}\sin \Delta m t$
term in the decay rate evolution.}} $B_{L}-B_{H}$
lifetime difference, the proper time evolution of the decay rate
can be written as
[3]
$$\Gamma (B^{0}(t)\rightarrow f_{i}) \;\propto e^{-\Gamma t} \biggl(
1- \sin 2 \alpha_{i} \;\sin \Delta m t \biggr)\;,$$
$$\Gamma (\bar B^{0}(t)\rightarrow f_{i}) \;\propto e^{-\Gamma t} \biggl(
1+ \sin 2 \alpha_{i} \;\sin \Delta m t \biggr)\;,$$
with $\Delta m$ being the $B_{L}-B_{H}$ mass difference.
Here we have taken into account that [14]
\begin{eqnarray}
 \biggl| {A(\bar B^{0}\rightarrow f_{i})\over A(B^{0}\rightarrow f_{i})}
\biggr| &\approx &1 \;,\\
 \biggl| {q\over p} \biggr| &\approx &1\;,
\end{eqnarray}
where $p,q$ are the Pais-Treiman coefficients [15] defining the mass
eigenstates in terms
of the flavor eigenstates $B^{0},\bar B^{0}$. The CP-asymmetry results
from an interference between the two processes
$$B^{0}\rightarrow f_{i} \;\;{\rm and}\;\; B^{0}\rightarrow \bar B^{0}\rightarrow
f_{i}\;,$$
and includes CP-violating effects in both mixing ($\vert\Delta B\vert=2$)
and decays ($\vert\Delta B\vert=1$). 
The angles of the unitarity triangle
can be expressed as
\begin{eqnarray}
&& \sin 2\alpha_{i} =\eta^{CP}_{i} \; {\rm Im}\; \biggl[ {q\over p}\;
{\langle f_{i}\vert {\mathcal{L^{CP}}}\vert B^{0}\rangle
\over \langle f_{i} \vert {\mathcal{L}}\vert B^{0}\rangle } \biggr]
= -\eta^{CP}_{i} \; \sin (2\phi_{D_{i}} + \phi_{M})\;,
\end{eqnarray}
where  $\eta^{CP}_{i}$ denotes the CP-parity of the final state ($-1$ for
 $\psi   K_{s}$ and $\rho^0  K_{s}$, and $+1$ for $\pi^{+} \pi^{-}$) and
 $\phi_{D_{i}}, \phi_{M}$ are the weak phases entering the
 $b\rightarrow q{\bar q}Q$ decay and $B-\bar B$ mixing diagrams, respectively. 
Note that no hadronic uncertanties are involved in this formula
and, in the Standard Model, the $\alpha_{i}$ are functions of the CKM
matrix elements only. Even though one cannot predict the exact values of
these angles due to large uncertanties in the CKM matrix, eq.(6) allows us
to verify generic features of the CKM approach such as the existence of the
unitarity triangle. Should  $\alpha_{i}$ not add up to $180^{\circ}$,
the necessity for an alternative theory of CP-violation would be manifest.
Thus, B decays into CP-eigenstates provide a useful and precise
tool in the search for physics beyond the Standard Model.

\section{Implications of Spontaneous CP-violation for B Decays}
Let us now proceed to evaluating the angles $\{ \alpha_{i} \}$  in the
context of the NMSSM. This model includes the MSSM superfields along
with an extra singlet superfield $\hat N$ and was first introduced
to rectify the so called ``$\mu -problem$'' [16].
A list of relevant interactions can be found in Refs.[11] and [17].

We assume that the initial Lagrangian conserves CP and
it is only the vacuum that breaks it.
In the process of electroweak symmetry breaking the neutral Higgs
components develop the following VEV's:
\begin{eqnarray}
&&\langle H_1^0 \rangle =v_1,\;\langle H_2^0 \rangle=v_2 e^{i\rho},
\langle N \rangle=n e^{i\xi}\;. \nonumber
\end{eqnarray}
If $\rho$ and/or $\xi$ are not equal to an integer multiple of $\pi$,
CP symmetry is violated. Through various interactions
these complex phases will enter the mass matrices and interactions
of the matter fields leading to observable CP-violating effects.

To estimate the consequent asymmetries in B decays,
first we will have to find the weak phase $\phi_{M}$
coming from the $B-\bar B$ mixing diagram.
Following the line of Ref.[11],
we adopt the following super-CKM ansatz (a squark version of the CKM matrix)
\begin{eqnarray}
&&{\tilde V} \approx \pmatrix{1&O(\epsilon)&O(\epsilon^2) \cr
                     O(\epsilon)&1&O(\epsilon)\cr
                     O(\epsilon^2)&O(\epsilon)&1\cr}
\end{eqnarray}
Here $\epsilon$ is of the order of Cabibbo mixing angle $\theta_{C}$.
Note that all the entries are real since we are considering spontaneous
CP-violation.

The real part of the $B-\bar B$ 
mixing is dominated by the SM box and chargino super-box diagrams (Fig.1a,b).
For simplicity, we assume that the gluino is sufficiently heavy and its
contribution to the $B-\bar B$ mixing is negligible.{\footnote{As long as the
gluino contribution does not dominate the $B-\bar B$ mixing, the essential
results of this paper remain unchanged.}}
There are three major contributions to the imaginary part
of the $B-\bar B$ mixing coming from the CP-violating diagrams in
Fig.2a,b and c.
The box diagram with Higgs exchange (Fig.2a) involves a complex phase
in the top quark mass insertion (this phase, of course, can be
absorbed into the Higgs vertex by a phase redefinition of $t_{R}$).
However, this diagram is suppressed by a factor of $(m_{b}/m_{W})^2$ and
can safely be neglected. The diagrams in Fig.2b,c contain phases in
propagators of the superparticles. In the case of $K-\bar K$ system,
an analog of the diagram 2b is responsible for a nonzero value for
$Re\;\epsilon$ [11].
To estimate its effect for the $B-\bar B$ system, we can repeat
the $K-\bar K$ mixing analysis
with heavy squarks $m_{\tilde q}^2 \gg m_{\tilde W}^2$ [12]. Note that
the diagram 2c, which did not play any role for the $K-\bar K$ system,
can give a significant contribution to the imaginary part of the
$B-\bar B$ mixing.
The corresponding $\Delta B=2$ operator is given by
\begin{eqnarray}
&& O_{\Delta B_{q}=2} =(k_{q}+ e^{i\phi} l_{q} + e^{2i\phi} l_{q}' )\;
 \bar d \gamma^{\mu} P_{L} b\;
                    \bar d \gamma_{\mu} P_{L} b , 
\end{eqnarray}
where $k_{q}$, $e^{i\phi} l_{q} $ and $e^{2i\phi} l_{q}'$ ($q=d,s$)
result from
the diagrams shown in Fig.1, Fig.2b and Fig.2c, respectively.
The weak phase  $\phi$ is a function of the complex phases of the Higgs VEV's
and is constrained to be between 0.01 and 0.1 (for the sake of definiteness, we
assume it to be positive) from the $K-\bar K$ and NEDM
analyses [11,12].

The Standard Model contribution is well known [19] and can be approximated by
\begin{eqnarray}
&&k_{q}^{SM} \approx {g^4\over 256 \pi^2 M_{W}^2}\; (V_{tb}V_{tq})^2\;.
\end{eqnarray}
Assuming that the first and second generation squarks are degenerate in mass
and the stop mass is different, we estimate the super-box contribution
(Fig. 1b) to be [18]
\begin{eqnarray}
&&k_{q}^{susy} \approx {g^4\over 192 \pi^2 m^2_{\tilde q }}\;
({\tilde V_{tb}}{\tilde V_{tq}})^2\;
\end{eqnarray}
with $m^2_{\tilde q }$ being the average squark mass.
The CP-violating super-box (Fig.2b) generates [12]
\begin{eqnarray}
&&  l_{q} \approx
{g^4 \over 128 \pi^2} \biggl({gm_{t}\over \sqrt{2} m_{W} sin\beta}  \biggr)^2 
\;\frac{v \;z\; m^2_{LR}}
{m_{\tilde W} m^4_{\tilde q } }  
\;
({\tilde V_{tb}}  {\tilde V_{tq}})^2 \;.
\end{eqnarray}
Here $z\sim 1$ is a partial cancellation factor [11];
 $v=\sqrt{v_1^2 + v_2^2}$;
 $m_{\tilde W}$ and $m_{t}$ denote the chargino
and top quark masses, respectively;
$m_{LR}$ is the left-right squark mixing, and $tan\beta=v_2 /v_1$.
Finally, the diagram in Fig.2c gives rise to
\begin{eqnarray}
&&  l_{q}' \approx
{g^4 \over 256 \pi^2} \biggl({gm_{t}\over \sqrt{2} m_{W} sin\beta}  \biggr)^4 
\;\frac{v^2 \;z'\; m^4_{LR}}
{m^8_{\tilde q } }  
\;
({\tilde V_{tb}}  {\tilde V_{tq}})^2 \;.
\end{eqnarray}
The factor $z'\sim 1$ results from the diagram
2c in which positions of $\tilde t_{L}$ and $\tilde t_{R}$ (as well as
$\tilde W$ and $\tilde H$) are interchanged. Such a diagram contributes with
opposite phase and leads to a partial cancellation.

To estimate a relative size of these couplings, let us assume a maximal
left-right mixing, $\tan\beta \sim 1$, $m_{\tilde W} \sim 100 GeV$ and
$m_{\tilde q }\sim 300 GeV$.
Using the super-CKM ansatz (7), it is not hard to see that
\begin{eqnarray}
&&l_{d}, l_{d}' \ll k_{d} \;,\\
&&l_{s}, l_{s}' \sim k_{s}\;.
\end{eqnarray}
An appreciable contribution of the CP-violating super-box to the
$B_{s}-\bar B_{s}$ mixing is an artifact of the chosen
super-CKM form (7). However, it reflects a general tendency for
models in which the mixing between the second and third generation squarks
is enhanced as compared to that of quarks.

As a result, the $O_{\Delta B_{s}=2}$ operator attains an overall
phase factor of $e^{i O(\phi )}$ and the corresponding weak phases are
\begin{eqnarray}
&& \phi_{M}(B_{s}) \sim \phi \;,\nonumber\\
&& \phi_{M}(B_{d}) \sim 0\;,
\end{eqnarray}
with $\phi \leq 0.1$.

Now we can proceed to calculating the remaining weak phase $\phi_{D_{i}}$.
In the Standard Model, the $b\rightarrow q{\bar q}Q$ decay is dominated
by the tree level process (Fig.3) and the weak phase results
from the complex CKM matrix elements entering the vertices. However, in our
case these entries are real. CP-violation must enter through a loop effect.
The simplest 1-loop diagram which involves complex phases in the propagators
of the superpartners is shown in Fig.4 (it is a version of the so called
``Superpenguin" diagram). Its $s\rightarrow q{\bar q}d$ analog
was calculated in [11,12] and shown to 
successfully describe the observed value of $\epsilon'$ in $K$ decays.
To get the corresponding 4-fermion effective interaction for the
 $b\rightarrow u{\bar u}d$ decay,
we simply need to change the super-CKM entries at the vertices. Then,
in the case $m_{\tilde q}^2 \gg m_{\tilde W}^2$, we obtain [12]
\begin{eqnarray}
&&O_{\Delta B=1}^{s.p.}= e^{i\phi}\;\vert f\vert\;
\bar d_{L} \gamma_{\mu} T^{a} b_{L}\;
        \bar q_{R} \gamma^{\mu} T^{a} q_{R}\;,  
\end{eqnarray}
with
\begin{eqnarray}
&&  \vert f\vert \leq
{g^2_3\; g^2 \over 576 \pi^2} \biggl({gm_{t}\over \sqrt{2} m_{W} sin\beta}  
\biggr)^2 
\;\frac{v \; m^2_{LR}}
{m_{\tilde q}^5 } \; 
\vert {\tilde V_{td}}{\tilde V_{tb}} \vert \;.
\end{eqnarray}
Here  $g_3, g$ are the strong and weak
couplings, respectively.
It is easy to see that for a reasonable choice of the parameters
($\tan \beta \sim 1,\; m_{\tilde q} \sim 300 GeV,\;m_{LR}/m_{\tilde q}
\leq 1$) the effective coupling $f$ is negligible as compared
to the Fermi constant which describes the tree level process in Fig.3.
The same argument equally applies to the decay mode
 $b\rightarrow c{\bar c}s$.
Hence, direct CP-violating effects in decay processes are strongly
suppressed and
the weak decay phases $\phi_{D_{1-3}}$ can be neglected. All CP-violation
in our scenario has to come from the $B-\bar B$ mixing and, consequently,
there is a universal phase which describes all CP-violating effects.
This is known as a {\it superweak} scheme of CP violation [20].

As a result, no CP-violation can be observed in $B_{d}-\bar
B_{d}$ systems. Eq.(6) now takes on the form
\begin{eqnarray}
&&\sin 2\alpha_{1}\approx 0 \;,\nonumber\\
&&\sin 2\alpha_{2}\approx 0 \;,\nonumber\\
&&\sin 2\alpha_{3}\approx \sin \phi \;.
\end{eqnarray}
Apparently,  the
$\{ \alpha_{i} \}$ fail to add up to $180^{\circ}$. However, the discrepancy
can be too small to be detected: since $\phi \sim 0.1$ the $\alpha_3$ is
no larger than a few degrees.

So far we have considered the implications of SCPV using a specific form
of the super-CKM matrix (7). With a more general super-CKM matrix,
one naturally distinguishes two possibilities:

1. The CP-asymmetries in $B,\bar B$ decays are negligible
leading to a flat unitarity triangle (this, for
example, happens when the super-CKM matrix duplicates the standard CKM
matrix). This is very different from the SM prediction
since, in the CKM model, all angles of the unitarity triangle
are typically larger than $10^{\circ}$ [21].

2. CP-violation in B decays is noticeable. Then some of the angles
$\alpha_{i}$ are measurably
different from zero (this requires a favorable super-CKM matrix, for example,
(7)). Since direct CP-violation is negligible in this model,
eq.(6) takes on the form
\begin{eqnarray}
&&\sin 2\alpha_{1}=\;\;\;\sin \phi_1 \;,\nonumber\\
&&\sin 2\alpha_{2}=-\sin \phi_1 \;,\nonumber\\
&&\sin 2\alpha_{3}=\;\;\;\sin \phi_2 \;
\end{eqnarray}
where $\phi_{1,2} \leq 0.1$. This case is represented
by a squashed ``triangle'' formed, for example, by $\phi_1 /2,\;
\pi\; -\;\phi_1 /2$, and $\phi_2 /2$. A deviation of
$\alpha_1 + \alpha_2 +\alpha_3$ from $180^{\circ}$ can be as large as
a few degrees. This is a direct contradiction to the Standard Model.
 
In both cases the deviations from the SM predictions are significant.
Note that the CP-asymmetries in this model are quite small.  That happens
because the same phase is responsible for both the NEDM and the CP-violating
effects in neutral meson systems. Since the EDM of individual quarks is
generated already at one loop level, the CP-violating phase has to be small
to comply with the experimental bound on the NEDM. As a result, large
CP-asymmetries cannot be accommodated within this model.

Another signature of spontaneous CP-violation may come from
independent measurements of the sides of the unitary triangle:
$\vert V(ub)\vert,\;\vert V(td)\vert$, and $\vert V(cb)\sin \theta_{c} \vert$.
Since CP-violation and quark mixing have different origins in SCPV models,
the relative values of these quantities do not have to be consistent
with the angles $\{ \alpha_{i} \}$.
In fact, $\vert V(ub)\vert,\;\vert V(td)\vert$,
and $\vert V(cb)\sin \theta_{c} \vert$ 
must form a completely flat triangle because all the CKM entries
have to be real. This observation combined with the constraints
on $\{ \alpha_{i} \}$ provides a very sensitive probe of the model.

To summarize, we have analyzed implications of
spontaneous CP-violation in the
simplest supersymmetric models for observable CP-asymmetries in B-decays.
We have argued that the SCPV approach realizes the {\it superweak}
scenario of CP-violation: all CP-violating effects are due to
$B-\bar B$ mixing. The expected asymmetries are significantly
smaller than those predicted by the Standard Model. A drastic
deviation from the SM predictions can, in principle, be observed
 in decays (1)-(3):
the corresponding CP-phases do not form the unitarity triangle
(this, however, would require quite precise experimental data).{\footnote{Large
CP-asymmetries in $B,\bar B$ decays ($\sin 2\alpha_1 \geq 0.4$) recently
reported by CDF collaboration [22] suggest that SCPV is not likely to be
the only source of CP-violation. However, the statistics at this time
does not preclude the scenario discussed in this paper.}}

Finally, it is worth mentioning that if both spontaneous and
(super-)CKM mechanisms of CP-violation are present then the
former can be responsible only for small corrections to the
SM values of the angles $\{ \alpha_{i} \}$. Since a deviation
of $\alpha_1 + \alpha_2 +\alpha_3$ from $180^{\circ}$ due to SCPV
does not exceed a few degrees, this model would be indistinguishable
from a susy model with general complex squark mixings [3].

To conclude, we see that a thorough study of B-phenomenology
can reveal the origin of CP-violation and shed light on the source
of new physics.

The author is grateful to Lay Nam Chang and Tatsu Takeuchi for discussions
and critical reading of the manuscript.

\newpage
{\bf Figure Captions}\\ \ \\
Fig. 1a,b    $\;\;\;\;$ Major contributions to the $B-\bar B$ mixing. \\ \ \\
Fig. 2a,b,c $\;$ Leading CP-violating contributions
           to the $B-\bar B$ mixing (all possible permutations are implied).\\ \ \\
Fig. 3    $\;\;\;\;\;\;\;\;\;$ Dominant contribution to $b\rightarrow q\bar q Q$. \\ \ \\
Fig. 4    $\;\;\;\;$ Leading CP-violating contribution to
          $b\rightarrow q\bar q Q$ (all possible positions of the left-right
          mixing are implied).

\end{document}